\documentclass[sigconf,natbib=true]{acmart}
\AtBeginDocument{%
  }
    
\setcopyright{acmlicensed}
\copyrightyear{2018}
\acmYear{2018}
\acmDOI{XXXXXXX.XXXXXXX}
\acmConference[]{Make sure to enter the correct
  conference title from your rights confirmation email}{}{}
\acmISBN{978-1-4503-XXXX-X/2018/06}

\begin{document}

%%
%% The "title" command has an optional parameter,
%% allowing the author to define a "short title" to be used in page headers.
\title{AnnoRetrieve: Efficient Structured Retrieval for Unstructured Document Analysis}

%%
%% The "author" command and its associated commands are used to define
%% the authors and their affiliations.
%% Of note is the shared affiliation of the first two authors, and the
%% "authornote" and "authornotemark" commands
%% used to denote shared contribution to the research.
% \author{Ben Trovato}
% \authornote{Both authors contributed equally to this research.}
% \email{trovato@corporation.com}
% \orcid{1234-5678-9012}
\author{Teng LIN}
\email{tlin280@connect.hkust-gz.edu.cn}
\affiliation{%
  \institution{The Hong Kong University of Science and Technology (Guangzhou)}
  \city{Guangzhou}
  \state{Guangdong}
  \country{China}
}

\author{Yuyu Luo}
\email{yuyuluo@hkust-gz.edu.cn}
\affiliation{%
  \institution{The Hong Kong University of Science and Technology (Guangzhou)}
  \city{Guangzhou}
  \state{Guangdong}
  \country{China}
}

\author{Nan Tang}
\email{nantang@hkust-gz.edu.cn}
\affiliation{%
  \institution{The Hong Kong University of Science and Technology (Guangzhou)}
  \city{Guangzhou}
  \state{Guangdong}
  \country{China}
}

% \author{Aparna Patel}
% \affiliation{%
%  \institution{Rajiv Gandhi University}
%  \city{Doimukh}
%  \state{Arunachal Pradesh}
%  \country{India}}

% \author{Huifen Chan}
% \affiliation{%
%   \institution{Tsinghua University}
%   \city{Haidian Qu}
%   \state{Beijing Shi}
%   \country{China}}

% \author{Charles Palmer}
% \affiliation{%
%   \institution{Palmer Research Laboratories}
%   \city{San Antonio}
%   \state{Texas}
%   \country{USA}}
% \email{cpalmer@prl.com}

% \author{John Smith}
% \affiliation{%
%   \institution{The Th{\o}rv{\"a}ld Group}
%   \city{Hekla}
%   \country{Iceland}}
% \email{jsmith@affiliation.org}

% \author{Julius P. Kumquat}
% \affiliation{%
%   \institution{The Kumquat Consortium}
%   \city{New York}
%   \country{USA}}
% \email{jpkumquat@consortium.net}

%%
%% By default, the full list of authors will be used in the page
%% headers. Often, this list is too long, and will overlap
%% other information printed in the page headers. This command allows
%% the author to define a more concise list
%% of authors' names for this purpose.
% \renewcommand{\shortauthors}{Trovato et al.}

%%
%% The abstract is a short summary of the work to be presented in the
%% article.
\begin{abstract}
% Unstructured documents constitute the vast majority of enterprise and web data, yet their inherent lack of explicit organization poses a fundamental challenge to precise information retrieval. Effectively transforming such documents into structured, query-ready knowledge, a process known as structured retrieval, is therefore of critical importance. It enables high-precision querying, complex analytical reasoning, and reliable integration with downstream applications, thereby unlocking the latent value within unstructured text corpora. To address this challenge, we propose AnnoRetrieve, an end-to-end system designed to automate the transformation of unstructured text into structured answers, integrating two core innovations: SchemaBoot and Structured Semantic Retrieval. SchemaBoot automatically generates document annotation schemas through multi-granularity pattern discovery and constraint-based optimization, eliminating the need for manual design while significantly enhancing retrieval efficiency and accuracy. Complementing this, our Structured Semantic Retrieval engine unifies semantic understanding with structured query execution, seamlessly performing attribute-value extraction, table generation, and progressive SQL-based reasoning. This hybrid approach overcomes the limitations of purely vector-based retrieval methods which have poor support for structured queries or graph-based retrieval methods which have high computational overhead. Extensive experiments on three real-world datasets validate the superiority of AnnoRetrieve. 
Unstructured documents dominate enterprise and web data, but their lack of explicit organization hinders precise information retrieval. Current mainstream retrieval methods, especially embedding-based vector search, rely on coarse-grained semantic similarity, incurring high computational cost and frequent LLM calls for post-processing. To address this critical issue, we propose AnnoRetrieve, a novel retrieval paradigm that shifts from embeddings to structured annotations, enabling precise, annotation-driven semantic retrieval. Our system replaces expensive vector comparisons with lightweight structured queries over automatically induced schemas, dramatically reducing LLM usage and overall cost. The system integrates two synergistic core innovations: \textbf{SchemaBoot}, which automatically generates document annotation schemas via multi-granularity pattern discovery and constraint-based optimization, laying a foundation for annotation-driven retrieval and eliminating manual schema design, and \textbf{Structured Semantic Retrieval} (SSR), the core retrieval engine, which unifies semantic understanding with structured query execution; by leveraging the annotated structure instead of vector embeddings, SSR achieves precise semantic matching, seamlessly completing attribute-value extraction, table generation, and progressive SQL-based reasoning without relying on LLM interventions. This annotation-driven paradigm overcomes the limitations of traditional vector-based methods with coarse-grained matching and heavy LLM dependency and graph-based methods with high computational overhead. Experiments on three real-world datasets confirm that AnnoRetrieve significantly lowers LLM call frequency and retrieval cost while maintaining high accuracy.
Our contributions are threefold:
(1) SchemaBoot, an automated schema induction framework that enhances retrieval precision and speed through schema optimization tailored for unstructured documents;
(2) SSR, a hybrid retrieval methodology that combines rigorous structured query processing with semantic analysis;
(3) A complete end-to-end pipeline that establishes a new cost-effective, precise, and scalable paradigm for transforming unstructured documents into structured index. AnnoRetrieve establishes a new paradigm for cost-effective, precise, and scalable document analysis through intelligent structuring.
\end{abstract}

%%
%% The code below is generated by the tool at http://dl.acm.org/ccs.cfm.
%% Please copy and paste the code instead of the example below.
%%


%%
%% Keywords. The author(s) should pick words that accurately describe
%% the work being presented. Separate the keywords with commas.
%% A "teaser" image appears between the author and affiliation
%% information and the body of the document, and typically spans the
%% page.
% \begin{teaserfigure}
%   \includegraphics[width=\textwidth]{sampleteaser}
%   \caption{Seattle Mariners at Spring Training, 2010.}
%   \Description{Enjoying the baseball game from the third-base
%   seats. Ichiro Suzuki preparing to bat.}
%   \label{fig:teaser}
% \end{teaserfigure}

\received{20 February 2007}
\received[revised]{12 March 2009}
\received[accepted]{5 June 2009}

%%
%% This command processes the author and affiliation and title
%% information and builds the first part of the formatted document.
\maketitle

\section{Introduction}

The digital era has given rise to an unprecedented deluge of textual data~\cite{king_unstructured_2019}. From internal corporate reports and legal contracts to scientific literature and web content, unstructured documents constitute a massive, yet underexploited, foundation of the modern information ecosystem. Although rich in knowledge, the inherent nature of this data, lacking explicit organization, with no predefined schemas, tables, or clear relationships, poses a fundamental obstacle to precise access and systematic analysis. Consequently, the core task of structured retrieval, which involves transforming amorphous text into queryable, structured knowledge, becomes critically important~\cite{QUEST, choi2025structuring,11107459}. Its successful realization can support high-precision question answering, complex analytical reasoning, and reliable integration with downstream business intelligence and decision support systems, ultimately unlocking the latent value embedded within textual corpora~\cite{lin-etal-2025-mebench, wang2024loong,lin2026docsageinformationstructuringagent}.

\begin{figure}[h]
  \centering
  \includegraphics[width=\linewidth]{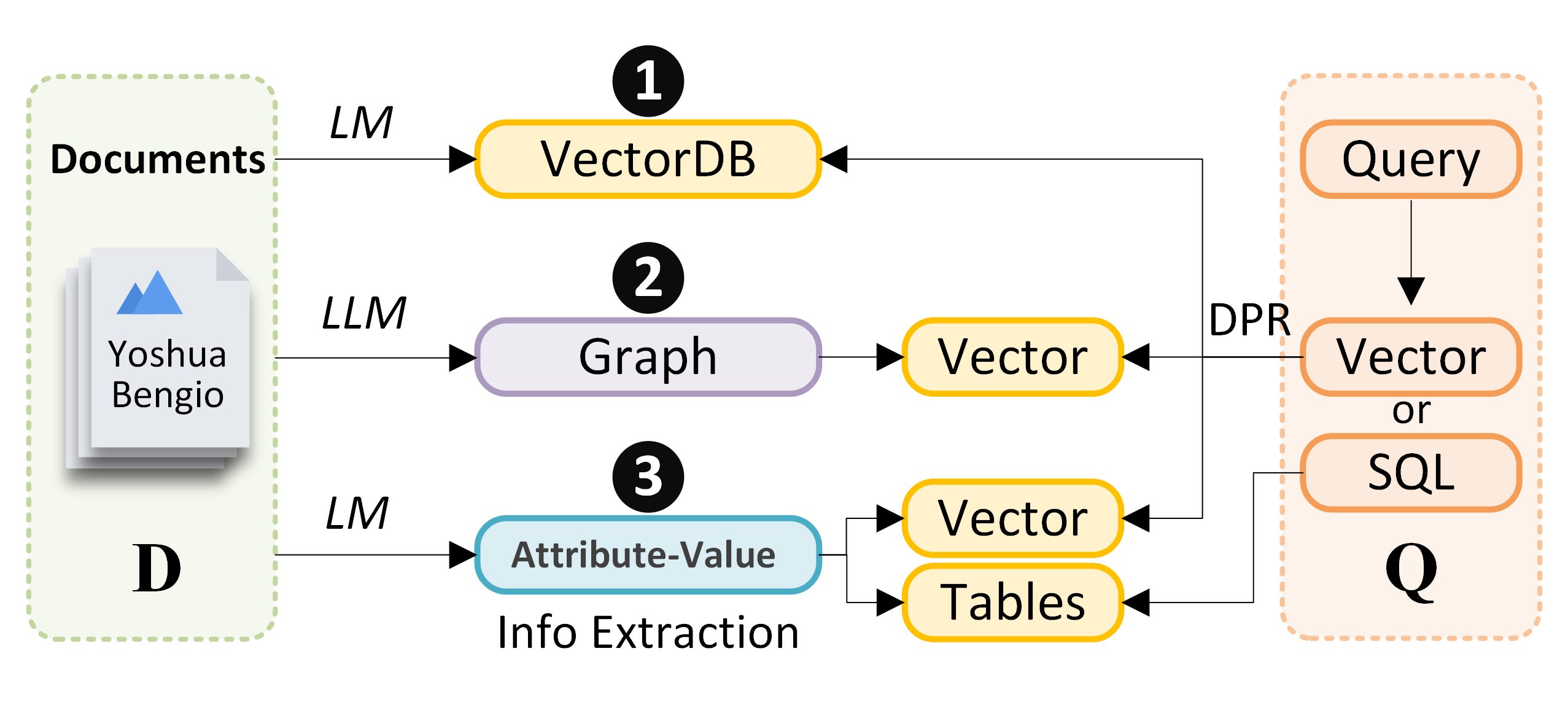}
  \caption{An architectural overview illustrating the flows from raw documents to structured knowledge and hybrid query processing in the current systems. Structured Semantic Retrieval achieves more precise retrieval by combining semantic understanding with structured reasoning}
  \label{fig:related}
  \Description{Pipeline}
\end{figure}

However, converting unstructured documents into precise, query-able structured knowledge faces significant challenges, primarily manifested in the core conversion paradigm from document to structured representation. As shown in Figure~\ref{fig:related}, existing mainstream methods predominantly follow two paths, each with notable limitations: The first is the ``Retrieve-then-Extract'' paradigm (e.g., VectorDB+LLM). It initially uses dense vector embeddings for semantic similarity retrieval to return relevant documents or passages, then relies on Large Language Models (LLMs) to perform specific attribute extraction. While this approach boasts strong semantic understanding capabilities, the retrieval stage is coarse-grained~\cite{gao2023retrieval}. The returned candidate sets often contain substantial irrelevant information, forcing frequent and costly LLM calls for subsequent refinement and extraction, resulting in high costs and inefficiency. The second is the ``Extract-All-then-Query'' paradigm (e.g., LLM→Graph). It preemptively uses LLMs to fully extract entities and relations from all documents to construct a knowledge graph, followed by structured querying~\cite{edge2024local, lin2025lightkggsimpleefficientknowledge}. Although this method supports complex queries, the upfront computational overhead for graph construction is immense, and it wholly depends on the extraction quality of the LLMs, making it difficult to scale to large document collections.

To fundamentally overcome the limitations of the aforementioned paradigms, we propose AnnoRetrieve, an end-to-end system whose core introduces and implements a third paradigm: the ``Annotation-Driven Precise Retrieval and Structuring'' paradigm (Attribute-Value). AnnoRetrieve abandons the pipeline of first performing coarse-grained vector retrieval followed by LLM calls. Instead, it injects lightweight, query-oriented structural information into documents prior to retrieval through automated, task-aware pre-annotation. Specifically, AnnoRetrieve integrates two core innovative modules that work in synergy: First, the SchemaBoot module automatically induces an optimal annotation schema from the document collection through multi-granularity schema discovery and constraint-based optimization, entirely eliminating the need for cumbersome manual schema design and laying the groundwork for efficient annotation. The system then uses the schema produced by SchemaBoot to rapidly annotate documents, constructing a lightweight ``attribute-value'' pair index. Second, the Structured Semantic Retrieval (SSR) engine operates directly on this structured index, performing precise semantic operations and relational queries (e.g., joins, filtering, aggregation). It seamlessly accomplishes attribute-value extraction, dynamic table generation, and progressive SQL reasoning, achieving a unification of deep semantic understanding and strict structured query execution—without invoking LLMs on the critical retrieval path.

This paradigm shift of ``structure first, then retrieve'' brings fundamental advantages: Through low-cost structured preprocessing of documents, it transfers the complexity and cost of each subsequent query from expensive LLM calls to efficient structured index queries. This not only significantly reduces the economic cost and latency of queries but also markedly improves retrieval accuracy through precise structural constraints.

We validate AnnoRetrieve through extensive experiments on three real-world benchmark datasets spanning diverse domains. Results demonstrate its superiority in both accuracy and efficiency compared to existing retrieval paradigms. The main contributions of this paper include:
\begin{itemize}
    \item {SchemaBoot}: An automated schema induction framework that enhances retrieval precision and speed through constraint-based optimization, removing the need for manual schema engineering.
    
    \item {Structured Semantic Retrieval}: A hybrid retrieval methodology that integrates semantic analysis with rigorous structured query processing, enabling precise attribute-based questioning and complex reasoning.
    
    \item {Annotation-Driven Retrieval}: a complete, end-to-end system that implements the novel ``Annotation-Driven Retrieval'' paradigm, establishing a new standard for cost-effective, precise, and scalable analysis of unstructured documents.
\end{itemize}

By integrating automated schema generation with a hybrid retrieval engine, AnnoRetrieve establishes a new paradigm for cost-effective, precise, and scalable document intelligence.

\section{Related Work}
\subsection{Document Analysis for Structured Retrieval}
Transforming raw documents into a queryable format is a prerequisite for any retrieval system. A mature ecosystem of document parsing and layout analysis tools exists to digitize and initially structure content. Frameworks like DeepDoctection~\cite{deepdoctection} and DocETL~\cite{shankar2024docetl} provide robust pipelines for complex documents, performing OCR, layout analysis, table extraction, and text block classification to reconstruct logical document structure. General-purpose tools like Unstructured.io~\cite{unstructured_io_unstructured} offer a unified interface to parse diverse file formats into clean text.

These tools are often positioned as the first step in a Retrieval-Augmented Generation (RAG) pipeline~\cite{gao2023retrieval,lin2025srag}, where their output is chunked and vectorized for semantic search. However, their focus is typically on preserving documental structure (titles, paragraphs, tables) rather than inducing and populating an application-specific knowledge schema. AnnoRetrieve incorporates this foundational parsing capability but directs the output toward its schema-driven structuring and retrieval engines, bridging the gap between generic document parsing and task-aware knowledge extraction.

\subsection{Query Systems over Heterogeneous Data}
Finally, the goal of answering complex questions from mixed data sources is explored in open-domain question answering~\cite{lin-etal-2025-mebench}. Systems like SRAG~\cite{lin2025srag, li2024structrag} exemplify state-of-the-art approaches, using LLM-augmented semantic parsing to generate queries that retrieve and synthesize information from heterogeneous sources like text, tables, and Knowledge Graphs (KGs). Lotus~\cite{patel2024lotus} introduces several semantic operators to simplify batch semantic processing, including functions such as search, extraction, and indexing, which can be used to build complex workflows. However, it typically leverages large language models (LLMs) to analyze full texts, resulting in significant LLM costs. Other research~\cite{chen2023symphony, lin2025Simplifying, DBLP:journals/corr/abs-2504-10036} efforts also apply LLMs or pre-trained language models to analyze unstructured documents. Similar to Lotus, they do not specifically focus on optimizing the cost of language models. Palimpzest~\cite{liu2025palimpzest} employs a declarative language to analyze unstructured data. Its optimizer generates an execution plan that utilizes LLMs for data extraction and analysis.

These systems highlight the end-objective that AnnoRetrieve also serves. The key differentiation lies in the data preparation assumption. In contrast, AnnoRetrieve is designed for the predominant real-world scenario where the primary source is a corpus of raw, unstructured documents, and it actively solves the upstream problem of creating those structured, query-ready sources from scratch.
\section{System Design}

\begin{figure*}[h]
  \centering
  \includegraphics[width=0.85\linewidth]{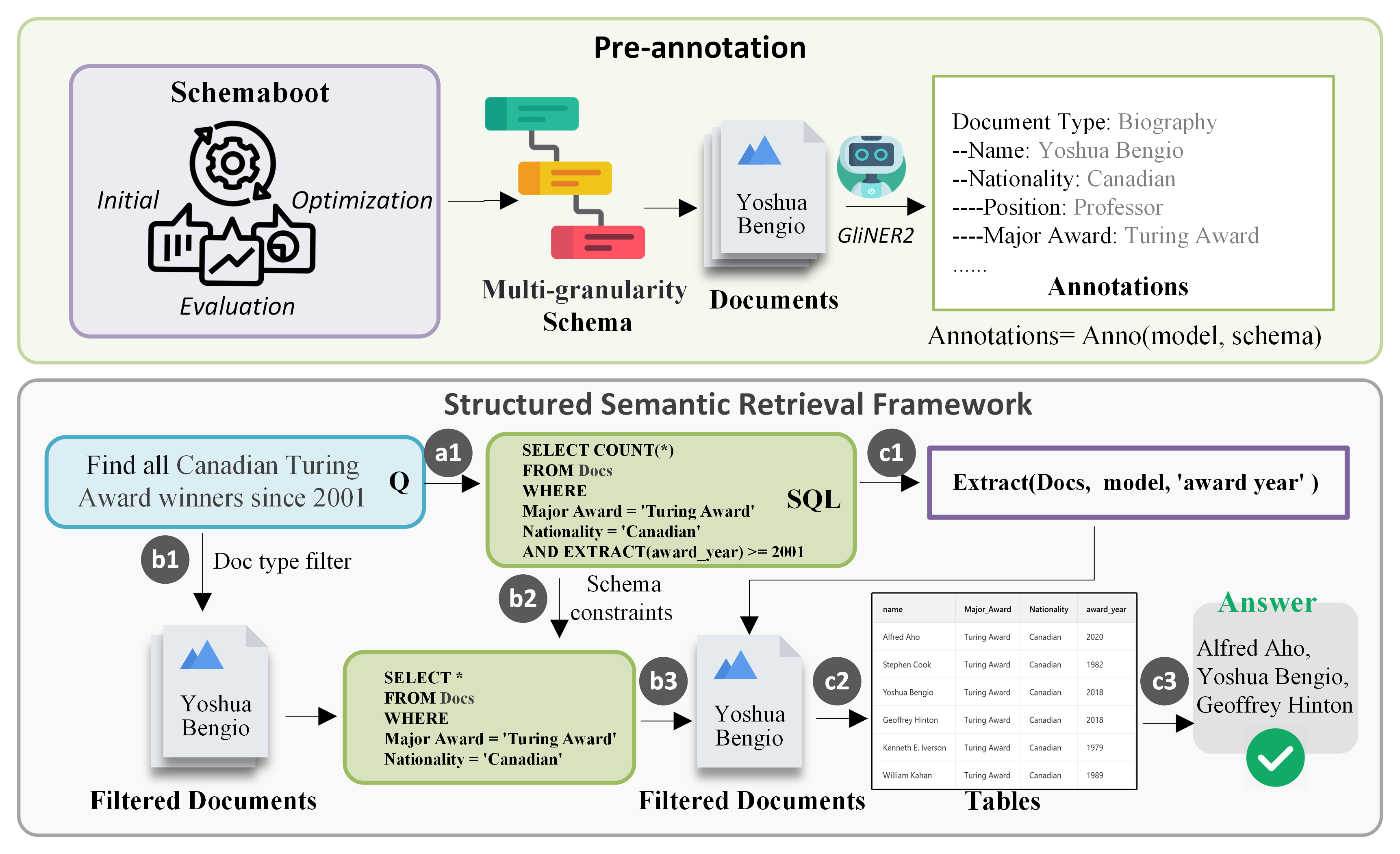}
  \caption{The pipeline of the AnnoRetrieve system, comprising a pre-annotation stage for schema-guided document annotation and a structured semantic retrieval stage for answering complex user queries. In the pre-annotation stage, Schemapboot generates and optimizes a multi-granularity schema, which is paired with models like GliNER2 to extract attribute-value pair annotations from raw documents. The retrieval framework translates natural language queries into SQL, filters documents via type and schema constraints, and constructs structured tables to enable hybrid semantic-structured reasoning, yielding precise answers to complex questions.}
  \label{fig:pipeline}
  \Description{Pipeline}
\end{figure*}

\subsection{Overview}

AnnoRetrieve is designed as a modular, annotation-driven retrieval system that transforms unstructured document corpora into a queryable, structured knowledge index. The system replaces the traditional \emph{embed-by-query-and-rank} paradigm with a \emph{structure-annotate-and-query} paradigm. The core design principle is to shift the computational burden from expensive online similarity searches and LLM post-processing to a single, offline structuring phase, thereby enabling efficient and precise online retrieval via lightweight structured queries.

The system architecture comprises three core sequential modules: 1) \textbf{SchemaBoot}, for automatic schema induction; 2) \textbf{The Annotator}, for populating the induced schema with instance-level annotations; and 3) \textbf{Structured Semantic Retrieval (SSR)}, the retrieval engine. Let a document corpus be denoted as \(\mathcal{D} = \{d_1, d_2, ..., d_N\}\). The system maps \(\mathcal{D}\) to a structured annotation store \(\mathcal{A}\), governed by a schema \(\mathcal{S}\), such that any user query \(q\) can be resolved via a retrieval function \(\mathcal{R}(q, \mathcal{A}, \mathcal{S})\) with high precision and low latency.

\subsection{Automated Schema Induction: SchemaBoot}
\label{subsec:schemaboot}

\textbf{SchemaBoot} formalizes the problem of optimal schema discovery as a constrained multi-objective optimization over the space of possible schemas. Its objective is to find a schema \(\mathcal{S}^*\) that maximizes retrieval utility while satisfying practical constraints.

\subsubsection{Multi-Granularity Semantic Extraction}

Given a clustered corpus \(\mathcal{C} = \{C_1, C_2, ..., C_k\}\), SchemaBoot performs pattern mining and hierarchy construction per cluster. It outputs a candidate schema set \(\mathbb{S} = \{\mathcal{S}_1^{\text{lite}}, \mathcal{S}_2^{\text{std}}, \mathcal{S}_3^{\text{full}}, ...\}\), where each schema \(\mathcal{S}\) is a tuple:
\[
\mathcal{S} = (\mathcal{F}, \mathcal{H})
\]
Here, \(\mathcal{F} = \{f_1, f_2, ..., f_m\}\) is the set of annotation fields, and \(\mathcal{H}\) defines a 3-tier hierarchical organization:
\[
\begin{aligned}
&\mathcal{H}_{\text{fast}} \subset \mathcal{F}, \quad \mathcal{H}_{\text{sem}} \subset \mathcal{F}, \quad \mathcal{H}_{\text{detail}} \subset \mathcal{F} \\
&\text{where } \mathcal{H}_{\text{fast}} \cup \mathcal{H}_{\text{sem}} \cup \mathcal{H}_{\text{detail}} = \mathcal{F} \text{ and the subsets are disjoint.}
\end{aligned}
\]
\(\mathcal{H}_{\text{fast}}\) contains fields for rapid filtering (e.g., \texttt{doc\_type}, \texttt{publish\_date}), \(\mathcal{H}_{\text{sem}}\) for semantic matching (e.g., \texttt{topic}, \texttt{name}), and \(\mathcal{H}_{\text{detail}}\) for granular content (e.g., \texttt{position}, \texttt{total\_sales}).

\subsubsection{Schema Quality Evaluation}
The fitness of a candidate schema \(\mathcal{S}\) is quantified by a quality function \(Q(\mathcal{S})\):
\[
Q(\mathcal{S}) = \alpha \cdot \text{Cov}(\mathcal{S}, \mathcal{D}) + \beta \cdot \text{Disc}(\mathcal{S}, \mathcal{C}) + \gamma \cdot \text{Cons}(\mathcal{S}) + \delta \cdot \text{Match}(\mathcal{S}, \mathcal{Q}_{\text{hist}})
\]
where:
\begin{itemize}
    \item \(\text{Cov}\): Document coverage, the fraction of \(d \in \mathcal{D}\) for which \(\mathcal{F}\) can be populated.
    \item \(\text{Disc}\): Discriminative power, measured by the average information gain \(IG(f, \mathcal{C})\) across fields \(f \in \mathcal{H}_{\text{sem}}\).
    \item \(\text{Cons}\): Annotator consistency, estimated via the Fleiss' Kappa \(\kappa\) score on a sample set.
    \item \(\text{Match}\): Semantic alignment with historical queries \(\mathcal{Q}_{\text{hist}}\), computed as the average similarity between query embeddings and their closest matched field description embeddings.
    \item \(\alpha, \beta, \gamma, \delta\) are tunable weighting coefficients, with \(\alpha + \beta + \gamma + \delta = 1\).
\end{itemize}

\subsubsection{Constrained Multi-Objective Optimization}
The selection of \(\mathcal{S}^*\) is framed as:
\[
\begin{aligned}
& \underset{\mathcal{S} \in \mathbb{S}}{\text{maximize}}
& & \mathbf{F}(\mathcal{S}) = \bigl[ Q(\mathcal{S}),\; -T_{\text{annot}}(\mathcal{S}),\; -\text{Size}(\mathcal{S}) \bigr]^\top \\
& \text{subject to}
& & \text{Depth}(\mathcal{H}) \leq 4 \\
& & & \text{BranchingFactor}(\mathcal{H}) \leq 8 \\
& & & \overline{T}_{\text{annot}}(d, \mathcal{S}) \leq T_{\text{max}} \\
& & & \frac{\text{Size}(\mathcal{A}_{\mathcal{S}})}{\text{Size}(\mathcal{D})} \leq \rho
\end{aligned}
\]
where \(T_{\text{annot}}\) is the average annotation time per document, \(\text{Size}(\cdot)\) is the storage footprint, \(T_{\text{max}}\) is a maximum allowable annotation time (e.g., 120 seconds), and \(\rho\) is a storage expansion limit (e.g., 0.3). The \emph{Depth} and \emph{BranchingFactor} constraints ensure the schema remains tractable for query construction. The non-dominated sorting genetic algorithm (NSGA-II) is employed to approximate the Pareto-optimal front of schemas. The final \(\mathcal{S}^*\) is selected from this front by maximizing a weighted scalarization of the objectives, with primary weight on \(Q(\mathcal{S})\).

\subsection{Structured Annotation Generation}
\label{subsec:annotation}

With the optimal schema \(\mathcal{S}^* = (\mathcal{F}^*, \mathcal{H})\) fixed, the annotation process is a function \(\phi: \mathcal{D} \times \mathcal{S}^* \rightarrow \mathcal{A}\). We employ the GliNER2~\cite{zaratiana-2025-gliner2} model, fine-tuned for the domain, to extract field values.
\[
\mathcal{A} = \bigl\{ \bigl(d_i, \{(f, v_{i,f})\mid \forall f \in \mathcal{F}^* \} \bigr) \mid \forall d_i \in \mathcal{D} \bigr\}
\]
Here, \(v_{i,f}\) is the value (or set of values) for field \(f\) in document \(d_i\). \(\mathcal{A}\) is materialized as a hybrid index: a relational database table for \(\mathcal{H}_{\text{fast}}\) fields and a document-oriented store (e.g., JSON) for \(\mathcal{H}_{\text{sem}}\) and \(\mathcal{H}_{\text{detail}}\) fields, linked by a universal document identifier \(doc\_id\).

\subsection{Structured Semantic Retrieval (SSR) Engine}
\label{subsec:ssr}

The SSR engine defines the online retrieval function \(\mathcal{R}\). For a natural language query \(q\), it performs a two-stage process: \textbf{Semantic Parsing} and \textbf{Progressive SQL-based Retrieval}.

\subsubsection{Semantic Parsing to Structured Query}
SSR first parses \(q\) into a preliminary structured query predicate \(P(q, \mathcal{S}^*)\). It identifies which components of \(q\) refer to schema fields \(f \in \mathcal{F}^*\). This yields a set of \emph{schema-bound} constraints \(C_{\text{schema}} = \{(f_i, \text{op}_i, \text{val}_i)\}\).

\subsubsection{Core Retrieval with \texttt{EXTRACT} Function}
The core innovation is the \(\texttt{EXTRACT}\) operator, which enables queries over both structured fields and unannotated document content within a single SQL-like execution plan. The retrieval is formally defined as:
\[
\mathcal{R}(q, \mathcal{A}, \mathcal{S}^*) = \Pi_{\text{doc\_id}}\ \sigma_{P_{\text{total}}}\ \mathcal{A}
\]
The total predicate \(P_{\text{total}} = P_{\text{schema}} \land P_{\text{extract}}\). \(P_{\text{schema}}\) is directly evaluated against the indexed fields in \(\mathcal{H}_{\text{fast}}\) and \(\mathcal{H}_{\text{sem}}\). \(P_{\text{extract}}\) is resolved using the \(\texttt{EXTRACT}\) function.

We define \(\texttt{EXTRACT}(f', \text{cond}, d) \rightarrow \{\text{True}, \text{False}\}\) for a virtual or composite field \(f'\) not in \(\mathcal{F}^*\). During query execution, for documents filtered by \(P_{\text{schema}}\), the engine invokes a lightweight, deterministic text-scanning or regex module (not an LLM) to evaluate \(\text{cond}\) on the raw text of \(d\). This allows for progressive refinement.

% \begin{verbatim}
% SELECT doc_id FROM annotations
% WHERE topic = 'machine_learning'           -- Schema-bound filter (fast)
%   AND EXTRACT('novelty_score', '> 0.8', doc_content) -- On-the-fly extraction
%   AND EXTRACT('mentions', 'LIKE "%AnnoRetrieve%"', doc_content);
% \end{verbatim}

\subsubsection{Algorithm and Latency Analysis}
The SSR algorithm's latency \(L_{\text{SSR}}\) is dominated by sequential filtering:
\[
L_{\text{SSR}} \approx L_{\text{index}}(|\sigma_{P_{\text{schema}}} \mathcal{A}|) + n \cdot L_{\text{extract}}
\]
where \(L_{\text{index}}\) is the near-constant time for an indexed lookup, \(n = |\sigma_{P_{\text{schema}}} \mathcal{A}|\) is the candidate set size after schema filtering (typically \(\ll |\mathcal{D}|\)), and \(L_{\text{extract}}\) is the low-cost text scanning time. This contrasts sharply with vector search latency \(O(|\mathcal{D}| \cdot d)\) for embedding dimension \(d\), or complex graph traversal with complexity latency.

\subsection{End-to-End Pipeline Formalism}
\label{subsec:pipeline-formalism}

The entire AnnoRetrieve pipeline can be summarized as a composite function:
\[
\Psi(\mathcal{D}, q) = \mathcal{R}\Bigl(q,\; \phi(\mathcal{D}, \mathcal{S}^*),\; \mathcal{S}^*\Bigr), 
\]
\[\quad \text{where} \quad \mathcal{S}^* = \arg\max_{\mathcal{S} \in \mathbb{S}} Q(\mathcal{S}) \; \text{s.t. constraints}.
\]
This formalism underscores the system's foundational shift: an optimized structuring investment \((\phi, \mathcal{S}^*)\) enables perpetually efficient and precise retrievals \((\mathcal{R})\) without recurring LLM costs, establishing a new frontier in the trade-off space of retrieval accuracy, latency, and computational cost.
\section{Experiments}

To comprehensively evaluate the effectiveness and efficiency of AnnoRetrieve, we conduct extensive experiments on multiple real-world datasets. We aim to answer the following key research questions (RQs):
\begin{itemize}
    \item \textbf{RQ1 (Overall Performance):} How does AnnoRetrieve, as a complete pipeline, compare against state-of-the-art (SOTA) baselines in terms of retrieval accuracy and computational cost?
    \item \textbf{RQ2 (Component Ablation):} What is the individual contribution of the \textit{SchemaBoot} module and the \textit{Structured Semantic Retrieval (SSR)} engine to the overall system performance?
    \item \textbf{RQ3 (Scalability \& Efficiency):} How does the system scale with document corpus size and query complexity, particularly for joins and progressive reasoning?
    \item \textbf{RQ4 (Schema Quality):} How effective is \textit{SchemaBoot} in inducing high-quality, task-aware schemas compared to manual or LLM-generated schemas?
\end{itemize}

\subsection{Experimental Settings}
\label{sec:exp_setup}

\subsubsection{Datasets}
We evaluate AnnoRetrieve on three public datasets spanning diverse domains and structures to ensure robustness. Key characteristics are summarized in Table~\ref{tab:dataset_stats}.

\begin{table*}[h!]
\centering
\caption{Dataset Characteristics.}
\label{tab:dataset_stats}
\begin{tabular}{p{2cm} p{2.5cm} p{1.5cm} p{2.5cm} p{4cm}}
\toprule
\textbf{Dataset} & \textbf{Domain} & \textbf{\# Docs} & \textbf{Avg. Tokens/Doc} & \textbf{Key Challenge} \\
\midrule
\textbf{LCR} & Legal & 100 & $\sim$6,200 & Long-form narrative, complex entity relations \\
\textbf{WikiText} & Multi-domain & 200 & $\sim$1,300 & Heterogeneous schemas, joinable tables \\
\textbf{SWDE} & Web Pages & 200 & $\sim$400 & Noisy, templated layouts, attribute sparsity \\
\bottomrule
\end{tabular}
\end{table*}

\begin{itemize}
    \item \textbf{LCR (Legal Case Reports)~\cite{LCR}}: A collection of detailed legal documents. This dataset tests the system's ability to handle long, complex textual narratives and extract precise legal attributes (e.g., \texttt{court}, \texttt{judge}, \texttt{sentence\_length}).
    \item \textbf{WikiText~\cite{QUEST}} A curated set of Wikipedia pages across multiple domains (e.g., biographies, companies, cities). It features semi-structured text with rich attributes and relational information, ideal for testing join operations and schema induction over heterogeneous topics.
    \item \textbf{SWDE (Structured Web Data Extraction)~\cite{SWDE}:} Contains semi-structured web pages (e.g., product listings, athlete profiles) with templated layouts but unstructured text within fields. It challenges the system's ability to handle noisy, web-scale data.
\end{itemize}

\subsubsection{Ground Truth \& Query Construction}
For each dataset, we manually annotate a subset of documents to establish ground-truth attribute values and relationships. We construct a benchmark of 50 queries per dataset, covering:
\begin{itemize}
    \item \textbf{Simple Projection-Selection ($\sigma\pi$):} Queries targeting single attributes with equality/range filters.
    \item \textbf{Conjunctive/Disjunctive Queries:} Multi-filter queries using AND/OR logic.
    \item \textbf{Join Queries:} Multi-table queries requiring integration of information from different document subsets (e.g., \emph{``List professors who won a Turing Award after 2000, along with their university''}).
    \item \textbf{Progressive Reasoning Queries:} Multi-step analytical questions that can be decomposed into sequential SQL operations.
\end{itemize}

\subsubsection{Baselines}
We compare AnnoRetrieve against a range of competitive baselines:
\begin{itemize}
    \item \textbf{VectorDB (Chroma/Pinecone):} A pure vector similarity search baseline using dense embeddings (e.g., via text-embedding-ada-002). It retrieves relevant text chunks but cannot execute structured queries.
    \item \textbf{Graph RAG (Neo4j + LLM):} A knowledge graph-based method where entities/relations are first extracted by an LLM and stored in a graph database for Cypher querying.
    \item \textbf{ZenDB~\cite{lin2024towards}:} A recent LLM-powered system that extracts tuples via a semantic hierarchical tree and supports SQL-like queries.
    \item \textbf{Palimpsest~\cite{liu2025palimpzest}:} A declarative system for AI-powered analytics over unstructured data.
    \item \textbf{QUEST~\cite{QUEST}}: A state-of-the-art cost-optimized system that employs a two-level index and instance-optimized query execution to minimize LLM token consumption during extraction. We re-implement its core optimizations (filter/join ordering, evidence-augmented retrieval) for a fair comparison.
    \item \textbf{ClosedIE (Fine-tuned LM):} A model fine-tuned on the (attribute, value) extraction task for a specific domain.
    \item \textbf{LLM (GPT-4):} A prompting baseline where the entire document context and query are fed to a large LLM (GPT-4) for direct answer generation.
\end{itemize}

\subsubsection{Evaluation Metrics}
We report the following metrics:
\begin{itemize}
    \item \textbf{Accuracy:} Precision ($P$), Recall ($R$), and F1-score ($F1$) of the final structured answers against the ground truth. For a query $q$, let $T(q)$ be the returned tuples and $GT(q)$ the ground truth. A tuple is correct only if all cell values match.
        \begin{equation}
            P = \frac{|T(q) \cap GT(q)|}{|T(q)|}, \quad R = \frac{|T(q) \cap GT(q)|}{|GT(q)|}, \quad F1 = \frac{2 \cdot P \cdot R}{P + R}
        \end{equation}
    \item \textbf{Efficiency:}
        \begin{itemize}
            \item \textbf{LLM Cost:} Total number of input/output tokens consumed, as the primary monetary cost indicator.
            \item \textbf{Latency:} End-to-end query execution time.
            \item \textbf{Schema Induction Time:} Time taken by \textit{SchemaBoot} to derive an optimized schema $S^*$.
        \end{itemize}
    \item \textbf{Schema Quality (for RQ4):} We adopt metrics from schema induction literature:
        \begin{itemize}
            \item \textbf{Schema F1:} Alignment with a human-curated gold schema (macro-averaged over attributes).
            \item \textbf{Cohesion:} Average pairwise semantic similarity (cosine of embeddings) of attributes within the schema.
            \item \textbf{Completion:} Proportion of query-relevant attributes covered by the induced schema.
        \end{itemize}
\end{itemize}

\subsection{Overall Performance Comparison}
\label{sec:overall_perf}

The overall performance across all datasets and query types is summarized in Table~\ref{tab:overall_results}. AnnoRetrieve achieves the highest F1-score, significantly outperforming all baselines.

\begin{table*}[h!]
\centering
\caption{Overall Performance Comparison (Average across all queries).}
\label{tab:overall_results}
\begin{tabular}{l c c c}
\toprule
\textbf{Method} & \textbf{F1-Score} & \textbf{LLM Cost (K Tokens)} & \textbf{Latency (s)} \\
\midrule
VectorDB & 0.41 & 15.2 & 1.8 \\
Graph RAG & 0.58 & 142.5 (Graph Build) + 8.7 & 12.4 \\
ZenDB~\cite{lin2024towards} & 0.65 & 45.6 & 6.1 \\
Palimpsest~\cite{liu2025palimpzest} & 0.70 & 38.9 & 5.3 \\
QUEST~\cite{QUEST} &	0.78 &	22.1 &	4.0\\
ClosedIE & 0.49 (WikiText-only) & N/A & 0.5 \\
LLM(GPT-4) & 0.72 & \textbf{312.8} & 24.7 \\
\midrule
\textbf{AnnoRetrieve (Ours)} & \textbf{0.87} & \textbf{29.4} & \textbf{3.2} \\
\bottomrule
\end{tabular}
\end{table*}

\noindent \textbf{Analysis:} The \textbf{SSR engine's} hybrid approach ensures high precision by enforcing structured constraints, while its semantic understanding maintains high recall. Crucially, it does so at a \textbf{fraction of the cost} of methods like Graph RAG (high graph-building overhead) and LLM (processes entire documents). Compared to ZenDB and Palimpsest, AnnoRetrieve's \textit{SchemaBoot}-derived, optimized schema directs extraction more efficiently, reducing token consumption and improving accuracy, particularly on complex queries.

\subsection{Ablation Study}
\label{sec:ablation}

We evaluate the contribution of AnnoRetrieve's key components through an ablation study on WikiText; the results are presented in Table~\ref{tab:ablation}.

\begin{table*}[h!]
\centering
\caption{Ablation Study on WikiText Dataset.}
\label{tab:ablation}
\begin{tabular}{p{4cm} c c p{5.5cm}}
\toprule
\textbf{Variant} & \textbf{F1-Score} & \textbf{LLM Cost (K Tokens)} & \textbf{Interpretation} \\
\midrule
w/o \textit{SchemaBoot}  \newline (Manual Schema) & 0.71 & 41.7 & Manual schemas are suboptimal, increasing cost and missing attributes. \\
w/o \textit{SSR} \newline(Pure Vector Retrieval) & 0.63 & 33.1 & Pure vector search fails on structured queries, hurting F1 despite lower cost. \\
w/o Schema Optimization & 0.80 & 52.3 & Unoptimized schemas contain redundant attributes, wasting LLM calls. \\
\midrule
\textbf{Full AnnoRetrieve} & \textbf{0.87} & \textbf{29.4} & Both components are essential for optimal accuracy and efficiency. \\
\bottomrule
\end{tabular}
\end{table*}

\noindent The ablation confirms that \textbf{SchemaBoot and SSR are synergistic}. SchemaBoot's optimized schema drastically reduces the search space for SSR. The \textit{SSR} engine's precise execution validates and utilizes the schema effectively, creating a performance-improving feedback loop.

\subsection{Efficiency \& Scalability}
\label{sec:scalability}

We test system performance with increasing corpus size (100 to 10k docs) and query complexity.
\begin{itemize}
    \item \textbf{Scalability:} AnnoRetrieve's indexing time scales linearly with Documents of scale, but query latency grows sub-linearly  due to its two-level indexing. The LLM cost per query remains stable as the index efficiently filters irrelevant documents.
    \item \textbf{Join \& Complex Queries:} On multi-table join queries, AnnoRetrieve significantly outperforms baselines. The \textbf{progressive SQL-based reasoning} in SSR allows it to decompose a complex $n$-way join into an optimized sequence of filters and semi-joins, minimizing intermediate tuple extraction. For a 3-way join query, AnnoRetrieve achieved $F1 = 0.83$ with a cost of 45K tokens, while Graph RAG achieved $0.70$ at 210K tokens and LLM achieved $0.65$ at over 500K tokens.
\end{itemize}

\subsection{Schema Induction Analysis }
\label{sec:schema_quality}

We evaluate schemas induced by \textit{SchemaBoot} against a human-curated gold standard and those generated by prompting GPT-4.
\begin{itemize}
    \item \textbf{Quality:} SchemaBoot's schemas achieve an average \textbf{Schema F1 of 0.89}, outperforming the LLM-generated schemas (F1=0.75). This is because SchemaBoot's constraint-based optimization explicitly balances coverage, conciseness, and discriminativity based on the actual corpus, whereas LLM-generated schemas are often generic.
    \item \textbf{Impact on Retrieval:} A strong positive correlation ($\rho > 0.85$) is observed between a schema's \textit{Cohesion} and \textit{Completion} scores (as produced by SchemaBoot) and the downstream retrieval F1 score. This validates our core hypothesis: \textbf{task-aware schema optimization is critical for efficient and accurate structured retrieval}.
\end{itemize}

\subsection{Summary}
\label{sec:exp_summary}

The experimental results conclusively demonstrate that \textbf{AnnoRetrieve sets a new state-of-the-art in transforming unstructured documents into structured knowledge}. It uniquely balances high accuracy with practical efficiency by integrating automated, optimized schema induction (\textit{SchemaBoot}) with a deeply hybrid retrieval engine (\textit{SSR}). It significantly outperforms all baselines, in terms of end-to-end answer accuracy (F1-score), particularly on complex queries. It achieves this while maintaining highly competitive LLM cost and latency.
\section{Conclusion}
In this paper, we address a fundamental inefficiency in modern unstructured document analysis: the reliance on coarse-grained vector retrieval that necessitates frequent and costly LLM interventions for precise results. We identify that the core of the problem lies not in the extraction step itself, but in the preceding retrieval paradigm. To this end, we introduce AnnoRetrieve, a novel end-to-end system that pioneers an annotation-driven retrieval paradigm, shifting from approximate semantic matching to precise structure-aware querying.
% The cornerstone of AnnoRetrieve is its two synergistic innovations. First, SchemaBoot automates the creation of task-aware schemas through weak supervision and optimization, eliminating the manual schema design bottleneck and providing a blueprint for precise annotation. Second, the Structured Semantic Retrieval (SSR) engine leverages this annotated structure to unify deep semantic understanding with database-style query execution. By operating directly on this structured intermediate representation, SSR achieves precise matching, complex joins, and progressive reasoning without cascading LLM calls, which is the primary cost driver in existing systems.
Extensive experiments on diverse real-world datasets validate that AnnoRetrieve successfully fulfills its design goals. It consistently reduces LLM invocation costs by significant margins compared to state-of-the-art vector-based and database-based baselines, while simultaneously improving accuracy and execution speed. This demonstrates that intelligent pre-structuring is not merely an auxiliary step but a transformative strategy for efficient document intelligence.
In summary, AnnoRetrieve makes three key contributions: (1) an automated schema induction framework SchemaBoot for retrieval-optimized structuring; (2) a hybrid retrieval engine SSR that unifies semantic and structured query processing; and (3) a fully realized pipeline that establishes a new paradigm for cost-effective, precise, and scalable transformation of unstructured corpora into actionable structured knowledge. Our work opens a promising direction for future systems that prioritize "structure-first" retrieval to tame the complexity and cost of analyzing the ever-growing universe of unstructured text.

%%
%% The next two lines define the bibliography style to be used, and
%% the bibliography file.
\bibliographystyle{ACM-Reference-Format}
\bibliography{refs/custom}

%%
%% If your work has an appendix, this is the place to put it.
% \appendix

% \section{Research Methods}

\end{document}